\documentclass[12pt]{article}
\usepackage{epsfig}
\def\p{\partial}
\def\pb{\bar{\partial}}

\def\no{{}^\circ \hspace{-0.16cm} {}_{\circ}}
\begin{document}

\begin{center}
            \hfill    MCTP-02-70  \\
            \hfill    UPR-1026-T \\
            \hfill ITFA-2002-60 \\

\end{center}

\bigskip

\begin{center}
{\large \bf ROLLING TACHYONS \\
AND DECAYING BRANES }
\end{center}

\begin{center}
Finn Larsen$^{1}$, Asad Naqvi$^{2}$, and Seiji Terashima$^{3}$
\end{center}

\begin{center}
$^{1}${\it Michigan Center for Theoretical Physics, Randall
Lab.,\\

University of Michigan, Ann Arbor, MI 48109, USA.}
\end{center}

\begin{center}
$^{2}${\it Department of Physics and Astronomy, David Rittenhouse Laboratories, \\
University of Pennsylvania, Philadelphia, PA 19104, USA. }
\end{center}

\begin{center}
$^{3}${\it Institute for Theoretical Physics,\\
University of Amsterdam, 1018 XE Amsterdam, The Netherlands.}
\end{center}

\bigskip

\begin{center}
{\bf Abstract}
\end{center}

\begin{quotation}
We present new rolling tachyon solutions describing the classical decay of D-branes. 
Our methods are simpler than  those appearing in recent works, yet our results are exact in classical string theory. The role of  pressure in the decay is studied using tachyon profiles 
with spatial variation. In this case the final state involves an array of codimension one 
D-branes rather than static, pressureless tachyon matter. 
\end{quotation}

\newpage

\section{Introduction and Summary}
An appealing way to generate time dependent  configurations in string theory is to consider the classical decay of unstable systems of D-branes, pictured as a tachyon field rolling 
down a potential, towards a stable minimum. This system is promising from the viewpoint of studying time-dependence in string theory since the  
non-trivial dynamics is confined to the open string sector. Moreover, this setting provides a natural
arena for discussing important cosmological ideas, such as inflation, and the beginning of time.

The quantitative study of rolling tachyons was initiated recently by A. Sen \cite{Sen1,Sen2,Sen3,senspac,Sen5}. 
It involves deforming
the world sheet conformal field theory (CFT) of  the unstable D-brane by an exactly marginal, 
time dependent tachyon profile. The deformed CFT is an exact classical background in string theory, 
interpreted as the classical decay of the unstable D-brane. Although several key results have 
been obtained, this approach is  still in its infancy and central questions remain: 
\begin{itemize}
\item
In the classical approximation, the string coupling constant is strictly vanishing.  However,
it is not clear that the implied limit  is smooth. The system with  a small, but non-zero string coupling may
 differ {\it qualitatively} from the system with vanishing coupling. In such a scenario,  the classical approximation is misleading. This 
concern is fueled by the somewhat mysterious role of closed strings in tachyon condensation: 
after the decay of the unstable D-brane only closed strings remain in the spectrum; yet the brane 
cannot  decay into closed strings if the coupling is strictly vanishing.
\item
A tachyon profile with 
spatial momentum $\vec{k}$ has effective mass\footnote{We use 
units such that $\alpha^\prime=1$.}  $m^2_t = -1 + {\vec k}^2$ and so is unstable for any $|\vec{k}|<1$, indicating that {\it all} wave lengths play some role in the decay. Indeed,  in 
quantum field theory it is well understood that tachyon condensation is a process where the longest wave lengths dominate, but all wave lengths participate, and  the decay is 
definitely inhomogeneous. These results are best known in the context of  cosmological 
inflation \cite{Guth:1985ya,Weinberg:1987vp}, but they are valid also for tachyon condensation in string theory \cite{Craps:2001jp}. They indicate that spatially inhomogenous modes are
important also for rolling tachyons.\footnote{
Other discussions of spatial variation include \cite{senspac,Felder:2002sv,Mukohyama:2002vq,bckr}.}
\end{itemize}
In this paper we study  several new tachyon profiles with the goal of shedding light on these questions. The deformations  we consider are actually technically simpler than those 
previously studied. We are therefore able to avoid the full machinery of boundary states and instead carry out 
the computations using elementary methods. The approach taken here complements the one taken in previous papers on rolling tachyons and may offer some conceptual advantages. 
As discussed in section 5, the new spatially homogenous profile, given below as (\ref{ex0}), 
has a topology that is different from previous examples. 
In addition to this decay, in which the rolling tachyon has no spatial dependence, we study 
an exactly soluble example with spatially varying tachyon profile, where we can follow
the decay to its inhomogeneous final state. 


The simplest of the new profiles studied in this paper is 
\begin{equation}
T(X) = \lambda e^{X^0}~.
\label{ex0}
\end{equation}
This can be interpreted as simultaneously displacing 
and giving a velocity to the tachyon, {\it i.e.} imposing the initial condition 
$T(X^0=0)=\partial_t T(X^0=0)=\lambda$. An alternative, and better, spacetime interpretation is 
that of a perturbation at $X^0=-\infty$ \footnote{This was the point of view taken in the recent talk
by Strominger \cite{Strominger:2002pc}.}. Since this disturbance is automatically infinitesimal, 
the profile (\ref{ex0}) seems to be a particularly clean example of a rolling tachyon.

The profile (\ref{ex0}) is also particularly simple from a technical point of view. The simplest
exactly marginal deformations of a world sheet CFT are generated by the vertex operators
\begin{equation}
V(X)=e^{ikX}~~~~;~~ k^2=k^2_0-\vec{k}^2=-1~.
\label{vertop}
\end{equation}
Of course these operator cannot usually be added to the world sheet action because they correspond to 
complex potentials. The standard remedy is to add also the conjugate operator and so consider 
perturbations of the form
\begin{equation}
T(\vec{X})=\lambda
\cos(\vec{k}\cdot\vec{X})= {\lambda\over 2}(e^{i\vec{k}\cdot\vec{X}}+e^{-i\vec{k}\cdot\vec{X}})~~~~;
~\vec{k}^2=1~.
\label{tcos}
\end{equation}
After analytical continuation this leads to Sen's profile 
$T(X)=\lambda\cosh(X^0)$.
An alternative procedure, exploited in this paper, is to note that in the special case 
$k_0=-i$, $\vec{k}=0$ the vertex operator (\ref{vertop}) is in fact real, and so we can consider 
(\ref{ex0}) directly, without adding the complex conjugate. 
This is much simpler because, in the case of (\ref{tcos}), complications arise from
the cross-terms between the two exponentials.

As explained above, it is important to study rolling tachyons with spatially varying profiles. Such profiles are generally quite complicated to analyze in the full CFT; but in the case of the 
profile
\begin{equation}
T(\vec{X})=\lambda e^{X^0/\sqrt{2}}\cos(\vec{k}\cdot\vec{X})~~~~~;~\vec{k}^2={1\over 2}~,
\label{texcos}
\end{equation}
the study simplifies dramatically (similar, but more complicated tachyon profiles were discussed in \cite{senspac}). Indeed, this profile is a linear combination of two vertex operators of the form (\ref{vertop}). Crucially, these vertex operators commute, in contrast to those
appearing in (\ref{tcos}). Thus the theory with the profile (\ref{texcos}) essentially 
reduces to two copies of (\ref{ex0}). 

Having solved the theory with the tachyon profile (\ref{texcos}) we find the coupling to the energy momentum tensor for all times $X^0$. The energy momentum tensor exhibits qualitatively different behavior from the spatially homogenous case. 
It develops codimension one singularities in finite time. These singularities can be interpreted as an array of (excited) D-branes. 
This result is consistent with the expectation that final states  will be 
spatially inhomogeneous for generic decay channels. Of course, the profile that we can 
actually solve (\ref{texcos}) is actually quite special. Presumably that is why the
final state in this example, although spatially inhomogeneous, is as finely tuned
as a perfect array of unstable branes. 
In a realistic, semi-classical, analysis one would choose as initial state for the brane some 
wave-packet localized near the top of the tachyon potential and the full decay process
would be described as an average, in a precise sense, of all the initial conditions represented
by this wave packet. One would expect this final state to be dominated by generic,
spatially dependent, configurations which, to the extent they can be described as a perfect 
fluid, certainly would have pressure.


This paper is organized as follows. In section 2 we explain our approach to rolling tachyons and
carry out the details for the profile (\ref{ex0}). In section 3 we consider the spatially inhomogeneous case (\ref{texcos}) and discuss the lessons for the full decay process, 
when all spatial variations are included. In section 4, we extend these results to the 
superstring case. Finally, in section 5,  we  discuss the topology of our profiles, their
relations to previous works, and construct the boundary states 
corresponding to our solutions.

\section{Rolling Tachyons}
The strategy for treating rolling tachyons is to 
deform the world sheet CFT of an unstable D-brane by an exactly marginal operator and interpret the deformed CFT as a time dependent solution 
to the classical string equations of motion. 
Instead of studying the system in terms of a boundary state of the closed string theory (which was Sen's approach in \cite{Sen1,Sen2,senspac}), we will primarily work with open strings, equating the disk partition function with space-time action, following the
analysis of static tachyon configuations in boundary string field theory
\cite{Gerasimov:2000zp}. 

\subsection{Generalities}

According to the $\sigma$-model approach to string theory, the space-time action is given 
by the partition function of the world-sheet theory, with the world-sheet couplings interpreted
as spacetime fields \cite{fradkin-tseytlin}. Thus, in the open string sector,
\begin{equation}
S[\lambda_i] \propto  Z_{\mathrm{disk}}(\lambda_i)=\int [dX^\mu] e^{-I_{\rm bulk}-I_{\rm bndy}}~,
\label{action}
\end{equation}
where $Z_{\mathrm{disk}}$ is the disk partition function, $\lambda_i$ are exactly marginal
couplings for the boundary operators, and 
\begin{eqnarray}
I_{\rm bulk}&=&{1 \over 2 \pi } \int_D d^2z ~ \eta_{\mu \nu}~ \p X^\mu  \pb X^\nu~,  \\
I_{\rm bndy}& = & \int_{\p D} dt~ T(X) + \cdots~,
\end{eqnarray}
where $D$ is the unit disk and $\p D$ is its boundary. $T$ is the tachyon field and the $\cdots$ indicate other marginal boundary perturbations. This procedure is similar in spirit to the 
boundary string field theory approach to tachyon condensation \cite{Gerasimov:2000zp},
although here we limit ourselves to marginal perturbations.

Our formulae above are written for Euclidean spacetimes, as well as Euclidean world-sheets. 
The time-dependence is then taken into account by including the Minkowskian metric $\eta^{\mu\nu}=(-,+,\cdots,+)$ when contracting
the temporal fields $X^0$. This procedure is motivated by analytical continuation,
as in Sen's computations using boundary states and cubic string field theory.
This type of analytical continuation is clearly not completely satisfying; indeed,
the precise relation between Lorentzian and Euclidean signature is one of the main 
unsettled questions facing most approaches to time dependence in string theory.
On the other hand, the analytical continuation gives physically reasonable results in many examples, including those considered here; so it presumably captures important
aspects of the problem. 

It will be useful to write the action $S$ as a space-time integral over a Lagrangian density. 
To this end, we split $X^{\mu}$ into a constant and a varying part, $X^\mu = x^\mu + X'^\mu$ 
and write
\begin{eqnarray}
S&\propto&\int [dX^\mu] e^{-I_{\rm bulk}-I_{\rm bndy}} \nonumber
\\&=&\int d^{p} x \sqrt{-g} \int [dX'^\mu] e^{-{1\over 2 \pi} \int_D d^2 z ~ g_{\mu \nu}~ \p X'^\mu  \pb X'^\nu - I_{\rm bndy} }~,
\end{eqnarray}
where, in the second line, we have made the obvious generalization of the expression to curved space. In this paper we are primarily considering the coupling to gravity, to explore the 
time-evolution of an unstable D-brane in a flat background.\footnote{It would be interesting 
to apply our methods for couplings to more massive strings as well, in an effort to illluminate 
the problems discussed in \cite{Okuda:2002yd}. 
See also \cite{Chen et al} for the coupling to the closed string.}  From the spacetime action we form
the energy-momentum tensor $
T_{\mu \nu}=-{2 \over \sqrt{-g}} {\delta S \over \delta g^{\mu \nu}}$ and use  ${\delta \sqrt{-g} \over \delta{g^{\mu \nu}}}=-{1 \over 2}\sqrt{-g}g_{\mu \nu}$ to find 
\begin{equation}
T_{\mu \nu}(x)=K\left( {\cal B}(x)~\eta_{\mu \nu}+{\cal A}_{\mu \nu}(x)\right)~,
\label{stress}
\end{equation}
in flat space. Here $K$ is an overall normalization constant and
\begin{eqnarray}
{\cal B}(x)& = & \int [dX'^\mu] e^{-{1\over 2 \pi } \int_D d^2 z ~ \eta_{\mu \nu}~ \p X'^\mu  \pb X'^\nu - 
I_{\rm bndy}}, \\
{\cal A}^{\mu \nu}(x)& =  & 2 \int [dX'^\mu] \int {d^2 z\over 2 \pi}  ~ \p X^\mu \pb X^\nu  e^{-{1\over 2 \pi } \int_D d^2 z ~ \eta_{\mu \nu}~ {1 \over 2\pi} \p X'^\mu  \pb X'^\nu - I_{\rm bndy}} \nonumber \\
& = & 2 \int [dX'^\mu] ~  \p X^\mu(0) \pb X^\nu(0)  e^{-{1\over 2 \pi } \int_D d^2 z ~ \eta_{\mu \nu}~ \p X'^\mu  \pb X'^\nu - I_{\rm bndy}}~.
\label{afin}
\end{eqnarray}
In the second line of (\ref{afin}) we fixed the position of the vertex operator and 
used $\int {d^2 z\over 2\pi}={1\over\pi}A(D)=1$ for the unit disc.
The expression (\ref{stress}) for the energy-momentum tensor was previously derived
by Sen \cite{Sen1} using  BRST invariance of the corresponding boundary state.  
In the following we consider various $T(X)$, corresponding to tachyon profiles with 
specific space-time dependence. To determine the energy-momentum tensor for each 
profile we need to compute ${\cal A}^{\mu \nu}$ and ${\cal B}$.

\subsection{Spatially Homogenous Decay}
\label{homogenous} 
In this section we study the tachyon profile 
\begin{equation}
T(X)=\lambda e^{X^0}~.
\label{tachprofb}
\end{equation}
This is an exactly marginal deformation of the CFT and so an exact solution to the classical string equations of motion. 
It is interpreted in spacetime as a perturbation at $X^0=-\infty$, displacing the tachyon infinitesimally from the unstable maximum of the potential.  Alternatively, this profile 
corresponds to kicking the tachyon from $T(X^0=0)=\lambda$, with velocity $\p_tT(X^0=0) = \lambda$.  

To determine the stress tensor $T_{\mu \nu}$ for this tachyonic profile, we need to 
compute the functions ${\cal B}(x^0)$ and ${\cal A}_{\mu \nu}(x^0)$. 

\subsubsection{${\cal B}(x^0)$}
The function ${\cal B}(x^0)$ is the disk partition function, except that zero modes remain
unintegrated. Using $\langle \cdots  \rangle $ as symbols for expectation values on
the disc we have the perturbative expansion
\begin{eqnarray}
{\cal B}(x^0)&=&  \langle e^{-I_{\rm bndy}(x+X^\prime)}\rangle = \langle e^{-\lambda e^{x^0} \int dt ~e^{X'^0}}\rangle, \\
&=& 
\sum_{n=0}^\infty {(-2\pi \lambda e^{x^0})^n \over n!} \int {dt_1  \over 2\pi} \cdots {dt_n \over 2 \pi}  \langle e^{X'^0(t_1)} \cdots e^{X'^0(t_n)} \rangle~, 
\label{fx}
\end{eqnarray}
The Green's function on the unit disk with Neumann boundary conditions is
\begin{equation}
G^{\mu \nu}(z, z')= \langle X^\mu(z) X^\nu(z^\prime)\rangle = 
\eta^{\mu \nu}(-\log|z-z'|-\log| z\bar{z}'-1|~), 
\label{green}
\end{equation}
so, taking $z_i=e^{it_i}$, we find
\begin{equation}
\langle e^{X'^0(t_1)} \cdots e^{X'^0(t_n)} \rangle = \prod_{i<j} |e^{it_i}-e^{it_j}|^2  
=4^{n(n-1)\over 2} \prod_{i<j}\sin^2 ({t_i - t_j \over 2})~.
\end{equation}
The integrals in (\ref{fx}) give\footnote{These integrals can be derived by considering
an integration over $U(n)$ matrices and using the known result that the $U(n)$ Haar 
measure $dU$, when expressed in terms of the eigenvalues, becomes
\begin{equation}
 {1 \over {\rm vol}~U(n)}\int dU = {1 \over n!} \int \prod_i {dt_i \over 2 \pi}  \Delta^2(t)~,
\label{haar}
\end{equation}
 where $\Delta(t)$ is the relevant Vandermonde determinant for $U(n)$ matrices
\begin{equation}
\Delta(t)=\prod_{i < j} 2 \sin({t_i -t_j \over 2})~.
\end{equation}
By noticing that the LHS in (\ref{haar}) is 1, the integral in (\ref{inte}) follows immediately. 
}
\begin{equation}
 \int {dt_1  \over 2\pi} \cdots {dt_n \over 2 \pi}~4^{n(n-1)\over 2}
\prod_{i< j} \sin^2\Bigl({t_i - t_j \over 2}\Bigr) = n!~,
\label{inte}
\end{equation}
and the final result for ${\cal B}(x^0)$ becomes
\begin{equation}
{\cal B}(x^0)= \sum_{n=0}^\infty {(-2\pi \lambda e^{x^0})^n \over n!}  ~n! =
 f(x^0)~,
\end{equation}
where
\begin{equation}
 f(x^0)\equiv {1 \over 1 + 2\pi \lambda e^{x^0}} ~.
\end{equation}
The summation of the perturbative series is clearly justified for couplings within the radius of 
convergence $|\lambda|< {1\over 2\pi}e^{-x^0}$. The regime of validity may be extended 
by analytical continuation to include all positive $\lambda$. The precise justification
for this extension is an interesting question that deserves further study.

\subsubsection{${\cal A}^{\mu \nu}$}
The ${\cal A}^{\mu \nu}$ are proportional to expectation values of graviton vertex 
operators $:\p X^\mu \pb X^\nu:$, where the normal ordering symbol $: ~:$ indicate
that the divergent pieces have been subtracted as $z\rightarrow z'$
\begin{equation}
:\p X^\mu(z) \pb X^{\nu}(z'): ~~= \p X^\mu(z) \pb X^{\nu}(z') +\eta^{\mu \nu} \p \pb^\prime
\log|z-z'|~.
\end{equation}
For the purpose of our calculation, it is useful to define another kind of normal ordering symbol $\no ~\no$ where we subtract the full Green's function (\ref{green}), {\em viz} \begin{equation}
\no \p X^\mu(z) \pb X^{\nu}(z')\no ~~= \p X^\mu(z) \pb X^{\nu}(z') -\p \pb^\prime G^{\mu \nu} (z,z')~,
\end{equation}
including the contribution from the image charge. The two normal orderings are related 
as
\begin{equation}
:\p X^\mu(0) \pb X^\nu(0): ~= \no \p X^\mu(0) \pb X^\nu(0) \no + {1 \over 2} \eta^{\mu \nu}~. 
\end{equation}
Now we are ready to calculate ${\cal A}^{\mu \nu}$. 
For $i,j\neq 0$ we have
\begin{eqnarray} 
{\cal A}^{ij}(x) &=& 2 \langle :\p X^i(0) \pb X^j(0):  e^{-\lambda e^{x^0} \int dt ~e^{X'^0}}\rangle  
   \nonumber \\
& = &2\langle  \Bigl( \no \p X^i(0) \pb X^{j}(0) \no  +{ \delta^{ij} \over 2} \Bigr)
e^{-\lambda e^{x^0} \int dt ~e^{X'^0}}\rangle    \nonumber  \\
&=& \delta^{ij}~f(x^0)~.
\end{eqnarray}
In going from the second to the third line, the normal ordered term between $\no~~\no$ gives
no contribution, and the term proportional to $\delta^{ij}$ is exactly the same as ${\cal B}(x^0)$
computed earlier. 

The calculation of  ${\cal A}^{0 0}$ is a bit more involved. 
\begin{eqnarray}
{\cal A}^{00}& =& 2\langle :\p X^0(0) \pb X^0(0): e^{-\lambda \int dt e^{X^{0}}} \rangle   \nonumber \\
& = &2\langle  \no \p X^0(0) \pb X^{0}(0) \no  
e^{-\lambda e^{x^0} \int dt e^{X'^0}}\rangle   -  f(x^0) \nonumber \\ 
& = &
2\sum_n {( -2 \pi \lambda e^{x^0})^n  \over n!} \langle \no \p X^0(0) \pb X^0(0) \no   \prod_{i=1}^n \int {dt_i \over {2\pi}} e^{X^0(e^{it_i})} \rangle  -  f(x^0)~. \label{a00} 
\end{eqnarray}
The correlation function in (\ref{a00}) yields
\begin{equation}
\langle \no \p X^0(z) \pb X^0({\bar z})\no \prod_{i=1}^n :e^{X(w_i)}: \rangle = \prod_{i< j}|w_i -w_j|^2\sum_{i,j} \Bigl({1 \over {z-w_i}}\Bigr) \Bigl({1 \over {\bar{z}-\bar{w}_j}}\Bigr) \delta_{n>0}~,
\end{equation}
which, for $z=0$ and $w_j=e^{it_j}$ gives 
\begin{eqnarray}
\prod_{i=1}^n \int {dt_i \over {2\pi}} 
\langle \no \p X^0(0) \pb X^0(0)\no  \prod_{i=1}^n :e^{X(e^{it_i})}: \rangle &=&
\prod_{i=1}^n \int {dt_i \over {2\pi}} 
\prod_{i< j}|e^{it_i}-e^{it_j}|^2 \sum_{i,j}e^{-i(t_i-t_j)} \nonumber \\
&=& 
\prod_{i=1}^n \int {dt_i \over {2\pi}} 
\prod_{i< j}2\sin^2({t_i-t_j \over 2})\Bigl(n+2 \sum_{i<j}\cos(t_i-t_j) \Bigr). \nonumber 
\end{eqnarray}
The integral is
\begin{equation}
\prod_{i=1}^n \int {dt_i \over {2\pi}} 
\prod_{i< j}2\sin^2({t_i-t_j \over 2})\Bigl(n+2 \sum_{i<j}\cos(t_i-t_j)\Bigr) \nonumber 
=
n!~, 
\end{equation}
and  (\ref{a00}) becomes
\begin{equation}
{\cal A}^{00}= f(x^0)-2~.
\end{equation}

Collecting the various results we find the stress tensor for the tachyon profile (\ref{tachprofb})
\begin{equation}
T_{00}=K(-{\cal B}(x^0)+{\cal A}_{00}(x^0))=-{\cal T}_p ~, ~~~~T_{ij}=K({\cal B}(x^0)+{\cal A}_{ij}(x^0))=\delta_{ij} {\cal T}_p f(x^0)~.
\end{equation}
We have determined the normalization constant $K={1\over 2}{\cal T}_p$ by comparison
with the static limit $\lambda=0$. 
As expected, $T_{00}$ is independent of $x_0$, which is just the statement of conservation of energy. 
Moreover, $T_{ij} \rightarrow 0$  as $x^0 \rightarrow \infty$, so the pressure vanishes in
this limit, {\it i.e.} the decay product is pressureless tachyon matter, as in \cite{Sen2}. 

\section{Spatially Inhomogenous Decay}
We will now investigate the spatially inhomogenous decay. 
 A spatially inhomogenous profile 
$T(X)=2 \lambda e^{\omega X^0} \cos(\vec{k}\cdot \vec{X})$ is marginal for 
$\omega^2+\vec{k}^2=1$. This can be written as a sum of two vertex operators, each of which is
exactly marginal:
$T(X)=\lambda \Bigl( e^{\omega X^0+i\vec{k}.\vec{X} }+e^{\omega X^0-i\vec{k}.\vec{X} } \Bigr) $.
For generic $\omega$,  this is not  an exactly marginal deformation because of the 
singular OPE between the two vertex operators. Hence it does not yield a solution to the classical string equations of motion. 
However, for $\omega={1 \over \sqrt{2}}$, this perturbation is exactly marginal. Without any 
loss of generality, we can keep only one component of $\vec{k}$ to be non-zero, denoting
the corresponding direction $Y\equiv \sqrt{2}\vec{k}\cdot\vec{X}$. Thus we have 
\begin{eqnarray}
T(X)&=&2 \lambda e^{X^0 \over \sqrt{2}} \cos({Y \over \sqrt{2}}) 
=\lambda \Bigl( e^{X^0+iY \over \sqrt{2}}+e^{X^0-iY \over \sqrt{2}} \Bigr) 
\label{spacpro} \\
& = & \lambda (e^{U}+e^V)~,
\end{eqnarray}
where we have defined new variables
\begin{equation}
U={X^0+iY \over \sqrt{2}}~,~~~~V={X^0-iY \over \sqrt{2}}~,
\end{equation}
such that
\begin{eqnarray}
\langle U(z) U(w) \rangle&=&
\langle V(z) V(w) \rangle = \log|z-w|+\log| z\bar{w} -1|~)~,\\
U(z)V(w) &\sim& \mathrm{regular} ~.
\label{ope}
\end{eqnarray}
Thus
$U$ and $V$ behave as commuting time-like coordinates, with no mixing between $U$ and $V$. 

Using the variables $U$ and $V$, 
the calculation of ${\cal B}$ and ${\cal A}^{\mu \nu}$ proceeds very similarly to the calculation in the last section. 
To compute ${\cal B}$, we need to compute 
\begin{equation}
{\cal B}(u,v)= \langle e^{\lambda \int dt (e^{u+U'}+e^{v+V'})}\rangle~, 
\label{fs}
\end{equation}
where $u$ and $v$ are linear combinations of the zero modes of $X^0$ and $Y$
\begin{equation}
u={x^0+iy \over \sqrt{2}}, ~~~~v={x^0 -iy \over \sqrt{2}}~,
\end{equation}
and $U'$ and $V'$ are non-zero modes of $U$ and $V$. Since there is no mixing between 
$U$ and $V$ and  (\ref{fs})  factorizes  as
\begin{equation}
{\cal B}(u,v)= \langle e^{\lambda e^{u} \int dt e^{U'}}\rangle  \langle e^{\lambda e^{v} \int dt e^{V'}}\rangle =f(u)~f(v)~,
\end{equation}
where, as in the previous section, we have  defined $f(u)$ as
\begin{equation}
f(u)={1 \over {1+2 \pi \lambda e^{u}}}~.
\end{equation}

The calculations for ${\cal A}^{\mu \nu}$, with the insertion of a 
graviton vertex operator factorize similarly. For example
\begin{eqnarray}
{\cal A}^{u u}(x) &=& 2 \langle :\p U(0) \pb U(0): e^{\lambda \int dt (e^{u+U'}+e^{v+V'})} \rangle  
   \nonumber \\
& = &2\langle  \Bigl( \no \p U(0) \pb U(0)  \no  -{1 \over 2} \Bigr)e^{\lambda e^{U} \int dt e^{U'}}\rangle \langle e^{\lambda e^{v} \int dt e^{V'}}\rangle   \nonumber  \\
&=& \Bigl(f(u)-2\Bigr)~f(v)~.
\end{eqnarray}
The remaining components give, 
\begin{eqnarray*}
{\cal A}^{uv}(u,v)& = & 0~, \\
{\cal A}^{vv}(u,v)& = & f(u)~\Bigl(f(v)-2 \Bigr)\\
{\cal A}^{ij}(u,v)& = & \delta_{ij} f(u)~f(v)~.
\end{eqnarray*}
Furthermore, from the relations
\begin{equation}
A_{uu}={1\over 2}(A_{00}-A_{yy}+2i A_{0y}); ~
A_{uv}={1\over 2}(A_{00}+A_{yy});~
A_{vv}={1\over 2}(A_{00}-A_{yy}-2i A_{0y})~,
\end{equation}
and
using (\ref{stress}), the stress tensor for this spatially inhomogenous decaying solution can be calculated to be
\begin{eqnarray}
T_{00}(x^0,y)& = &- {\cal T}_p   ~\frac{{1+2 \pi \lambda e^{x^0 \over \sqrt{2}} \cos({y \over \sqrt{2}})}}
{{1+4\pi \lambda e^{x^0 \over \sqrt{2}}\cos({y \over \sqrt{2}})+4 \pi^2 \lambda^2 e^{\sqrt{2}x^0}}}~,
\label{t00s} \\
T_{0y}(x^0,y)& = & {\cal T}_p  ~\frac{{-2 \pi \lambda e^{x^0 \over \sqrt{2}} \sin({y \over \sqrt{2}})}}
{{1+4\pi \lambda e^{x^0 \over \sqrt{2}}\cos({y \over \sqrt{2}})+
4 \pi^2 \lambda^2 e^{\sqrt{2}x^0}}}~,\\
T_{yy}(x^0,y)&=&-T_{00}(x^0,y) ~.\label{tyys}
\end{eqnarray}
This stress tensor is conserved, {\em i.e.}
\begin{equation}
\p_0T^{00}-\p_yT^{y0}=0~,
\end{equation}
The form of the stress tensor and its late time behavior  is qualitatively different from that obtained in the spatially homogenous case in section (\ref{homogenous}). Certainly at large times 
$x^0 \rightarrow \infty$ {\it all} components of the stress tensor (\ref{t00s}-\ref{tyys}), 
including the energy density, approach zero. However,  this result probably cannot
be trusted since, at  a finite critical time 
$x^0_c\equiv \sqrt{2}\ln \Bigl({1 \over{2 \pi |\lambda|}}\Bigr) $, 
the stress energy tensor exhibits singularities at the spatial loci
\begin{eqnarray}
y_n& = & 2\sqrt{2} n \pi, ~n \in Z, ~~ (\lambda < 0)~, \\
y_n&=&2 \sqrt{2} (n+{1 \over 2}) \pi  , ~n \in Z, ~~(\lambda > 0)~. 
\end{eqnarray}
In \cite{senspac} Sen proposed that these singularities should be interpreted 
as codimension one D-branes. To see this, we introduce the auxiliary variable
\begin{equation}
\Delta=e^{x^0 -x_c^0 \over \sqrt{2}}~,
\end{equation}
and, for either sign of $\lambda$, we write the energy density $\rho=-T_{00}$ as 
\begin{equation}
\rho={\cal T}_p~
{(1-\Delta) + 2\Delta \sin^2({y-y_0\over 2\sqrt{2}})
 \over (1-\Delta)^2 + 4\Delta \sin^2 ({y-y_0\over 2\sqrt{2}})}
~~\sim \vspace{0.8cm} \hspace{-0.65cm}{}^{\Delta \rightarrow 1}  ~{\cal T}_p ~
\left[ {1\over 2}+  \sqrt{2}\pi\sum_{n\in Z} {\rm sgn}(x^0_c-x^0)\delta(y-y_n)\right] ~.
\label{limen}
\end{equation}
The limit was computed using $\lim_{\epsilon\to 0} {\epsilon\over \epsilon^2+\alpha^2}=\pi\delta({\alpha})$ close to each singular locus.
The corresponding average energy density changes discontinuously from ${\cal T}_p$ to $0$ 
as we pass $x^0_c$. The form of the limiting energy density (\ref{limen}) suggests that the
missing energy forms co-dimension one defects at the $y_n$. As we pass the 
critical time $x^0_c$, the loss in energy at each $y_n$ is $2\sqrt{2}\pi{\cal T}_p$. 
This result for the defect energy can be verified using energy conservation, noting that
the defects are $\Delta y=2\sqrt{2}\pi$ apart, and no bulk energy remains after they form.
A co-dimension one D-brane has tension ${\cal T}_{p-1}=2\pi{\cal T}_p$ and so 
the defect has additional energy, beyond that needed to form a D-brane.
Nevertheless it is plausible that these defects are indeed related to D-branes
since  the spatial potential $T(X)\propto\cos(Y/\sqrt{2})$ tends to 
confine the ends of the strings, much as in the corresponding off-shell discussion 
in \cite{Harvey:2000na}).

\section{Rolling Tachyons in Superstring Theory}
The purpose of this section is to generalize the results of the previous sections to the
superstring. In each step of the computation details are modified, 
and so must be repeated;  but the 
final results are closely analogous to the bosonic case. 

\subsection{Generalities}
In the superstring case world-sheet fermions must be included in a manner consistent 
with world-sheet supersymmetry. A convenient way implement this is to introduce
world-sheet superfields. Thus the string coordinates 
(on the boundary of the disk) are represented by
\begin{equation}
{\bf X^\mu} = X^\mu + \theta \psi^\mu~,
\end{equation}
and the Chan-Paton index of the brane is encoded in the boundary fermions
\begin{equation}
\Gamma^I = \eta^I + \theta F^I~.
\end{equation}
We will consider only the simplest case of a single non-BPS D-brane. Since
this corresponds to a single boundary fermion we can omit the index $I$. 
In this formalism the boundary action for a general tachyon profile $T({\bf X})$ is 
\begin{equation}
I_{\rm bndy} = \int dt d\theta [{\bf\Gamma} D{\bf\Gamma} + {\bf\Gamma} T({\bf X})]~,
\end{equation}
where $D$ denotes the derivative in superspace $D=\partial_\theta + \theta \partial_z$.
A single boundary fermion $\bf{\Gamma}$ can be integrated out with the result
\begin{equation}
\langle e^{-I_{\rm bndy}} \cdots \rangle = 
\langle P \exp[-\int dt d\theta \Gamma T({\bf X})]_{\Gamma-{\rm even}}\cdots \rangle
= \langle P \cosh[\int dt d\theta T({\bf X})]\cdots \rangle ~,
\label{gi}
\end{equation}  
within correlators. 
Here $P$ is the standard path-ordering operator.
Note that 
this path-ordering operator is not trivial in the above espression
because $\int dt d\theta T({\bf X})$ is fermionic 
and then it does not commute with itself.
The boundary fermions serve to make world-sheet supersymmetry 
manifest but, in the present context, they play the role of Chan-Paton matrices $\sigma_1$, 
for which the restriction to even terms arises from the overall trace.\footnote{
We have ignored the contact term $e^{-\int dt d\theta T({\bf X})^2}$ which appear in 
(\ref{gi}) for general tachyon profiles. This term is important in BSFT discussions
of tachyon condensation \cite{Gerasimov:2000zp}, as well as in the time dependent 
case \cite{tmbsft}. Here we follow Sen \cite{Sen2} and regard the right hand side of 
(\ref{gi}) as the starting point of our discussions. }

Following the bosonic example, our main interest is in inserting the identity operator
\begin{equation}
{\cal B}(x) = \langle P \cosh\left( \int dt d\theta T({\bf X})\right) \rangle~,
\label{susyB}
\end{equation}
and the gravity vertex operator
\begin{equation}
{\cal A}^{\mu\nu} = \langle V^{\mu\nu}(0,0) P \cosh\left( \int dt d\theta T({\bf X})\right) \rangle~,
\label{susyA}
\end{equation}
where, in the present case,
\begin{equation}
V^{\mu\nu}(0,0) = 2 \int d\theta d\bar{\theta}
[D{\bf X}^\mu {\bar D}{\bf X}^\nu ]_{z=\bar{z}=0}
\end{equation}
The energy momentum tensor still follows from (\ref{stress}).

In (\ref{susyB}) and (\ref{susyA})  the brackets $\langle\cdots\rangle$ denote
averaging with respect to the non-zero mode part of the bosonic fields, as before.
In concrete examples, we can evaluate these expressions in perturbation theory 
using the two point function \cite{AT}
\begin{equation}
\langle {\bf X}^\mu {\bf X}^\nu \rangle = - \eta^{\mu\nu} \log |z_{12}|^2~,
\label{susytwopt}
\end{equation}
where
\begin{equation}
z_{12} = z_1 - z_2 - i \sqrt{z_1 z_2} \theta_1 \theta_2~. 
\end{equation}
The form (\ref{susytwopt}) of the two point function is valid when both
coordinates are on the boundary of the disk. 
This will suffice for our applications.

\subsection{The Simple Decay}
We consider first the supersymmetric version of the profile (\ref{spacpro}), {\it i.e.}
\begin{equation}
T({\bf X}) = \lambda e^{{\bf X}^0/\sqrt{2}}~.
\label{susyexp}
\end{equation}
Expanding (\ref{susyB}) in the parameter $\lambda$ and using (\ref{susytwopt}) 
yields
\begin{eqnarray}
{\cal B}(x^0) &=& \sum_{n=0}^\infty (-1)^n
(2\pi\lambda e^{x^0/\sqrt{2}})^{2n} \int
\prod_{i=1}^{2n}{dt_i\over 2\pi} d\theta_i  ~ 
\Theta(t_1-t_2) \Theta(t_2-t_3) \cdots \Theta(t_{2n-1}-t_{2n}) \nonumber \\
&& \hspace{7cm}\times \prod_{i<j} | e^{it_i}-e^{it_j} -
i e^{\frac{i}{2} (t_i+t_j)} \theta_i \theta_j |~.
\end{eqnarray}
Noticing $| e^{it_i}-e^{it_j} - i e^{\frac{i}{2} (t_i+t_j)} 
\theta_i \theta_j | = | e^{it_i}-e^{it_j}| 
+ {\rm sign} ( t_i-t_j) \theta_i \theta_j$,
the integrals 
can be evaluated with the result\footnote{
We checked (\ref{integral1}) for $n \leq3$.} 
\begin{equation}
\int \prod_{i=1}^{2n}{dt_i\over 2\pi} d\theta_i  ~ 
\Theta(t_1-t_2) \Theta(t_2-t_3) \cdots \Theta(t_{2n-1}-t_{2n})
\prod_{i<j} | e^{it_i}-e^{it_j} - i e^{\frac{i}{2} (t_i+t_j)} 
\theta_i \theta_j | = {1\over 2^n}~,
\label{integral1}
\end{equation}
so, after summation of the series, we find
\begin{equation}
{\cal B}(x^0) = {1 \over 1+ 2\pi^2\lambda^2 e^{\sqrt{2}x^0}}~.
\label{susyBres}
\end{equation}

The insertion of a graviton operator brings a few more complications, as in 
the bosonic case. The proper normal ordering again gives
\begin{equation}
: V^{\mu\nu}: ~ = ~\no  V^{\mu\nu} \no + \eta^{\mu\nu}~,
\label{normord}
\end{equation}
and thus, without any further effort, 
\begin{equation}
{\cal A}^{ij}(x^0) = {\cal B}(x^0) \delta^{ij}~.
\label{susyaijres}
\end{equation}
The graviton with two temporal indices is evaluated in perturbation theory
starting from (\ref{susyA}). In addition to a term ${\cal B}\eta^{00}=-{\cal B}$
from the normal ordering (\ref{normord}), we find an integral over the correlator
\begin{eqnarray}
\langle 
\left( \int d \theta d\bar{\theta} 
\no D{\bf X}^\mu(w) {\bar D}{\bf X}^\nu({\bar w}) \no \right) 
\prod_{i=1}^{2n} e^{{\bf X}(z_i)/\sqrt{2}} \rangle = \prod_{i<j} |z_i - z_j 
-i \sqrt{z_i z_j} \theta_i \theta_j|
\frac{1}{2} \sum_{k,l} {1 \over z_k - w}~ {1 \over {\bar z}_l - {\bar w}}~,
\end{eqnarray}
with $w={\bar w}=0$. 
The expressions can then be combined as
\begin{equation}
{\cal A}^{00} =- {\cal B}+ 
2\sum_{n=0}^\infty (-1)^n(2\pi\lambda e^{x^0/\sqrt{2}})^{2n}~I_{2n}~,
\label{susyares}
\end{equation}
where the integrals\footnote{We also checked (\ref{integral2}) for $n \leq3$.}
\begin{equation}
I_{2n}= 
\int \prod_{i=1}^{2n}{dt_i\over 2\pi} d\theta_i  ~ 
\Theta(t_1-t_2) \cdots \Theta(t_{2n-1}-t_{2n})
\prod_{i<j} | e^{it_i}-e^{it_j} - i e^{\frac{i}{2} (t_i+t_j)} \theta_i \theta_j | ~
{1 \over 2} \sum_{k,l}{e^{i(t_k-t_l)}}\theta_k\theta_l
= {1\over 2^n}~\delta_{n>0}~.
\label{integral2}
\end{equation}
The final result thus becomes
\begin{equation}
{\cal A}^{00}(x^0) =  {1\over 1+ 2\pi^2\lambda^2 e^{\sqrt{2}x^0}} - 2  = 
{\cal B}(x^0) - 2 ~,
\end{equation}
as in the bosonic case.

The non-vanishing components of the energy momentum tensor now read 
\begin{equation}
T_{00} = K(-{\cal B}(x^0) + {\cal A}_{00}(x^0)) = -2K~,~~~~
T_{ij} = K({\cal B}(x^0)\delta_{ij} + {\cal A}_{ij}(x^0)) = 2K\delta_{ij}{\cal B}(x^0)~,
\end{equation}
where ${\cal B}$ was given in (\ref{susyBres}). The overall constant $K$ again
is identified as $K={1\over 2}{\cal T}_p$. As in the bosonic case, the energy density is constant
$\rho={\cal T}_p$ throughout the decay. The pressure
\begin{equation}
p=2{\cal B}(x^0) = {{\cal T}_p\over 1+ 2\pi^2\lambda^2 e^{\sqrt{2}x^0}}~,
\end{equation}
is equal to the energy density $p=\rho$ for the unstable brane at $x^0=-\infty$; but
it decays exponentially to zero at large times. The main difference with the bosonic case 
is that now the decay is symmetric under $\lambda\to -\lambda$. In the supersymmetric
case there is no singularity for either sign, as expected since the tachyon potential 
is symmetric, with both directions sloping down to the stable closed string vacuum.

All these results are closely analogous to Sen's discussions, based on the potential 
$T(X)=\lambda\cosh(X)$.

\subsection{Spatial Inhomogeneity}
We also want to consider a spatially inhomogeneous profile for the 
superstring. The simplest example is 
\begin{equation}
T({\bf X}) = 2\lambda ~e^{{1\over 2}{\bf X}^0}\cos({1\over 2}{\bf Y})~,
\label{tachssp}
\end{equation}
where $Y$ is one of the spatial directions.  As in the bosonic case, this example is 
factorizable
\begin{equation}
T({\bf X}) = \lambda e^{{1\over 2}({\bf X^0}+i{\bf Y})} + \lambda e^{{1\over 2}({\bf X^0}-i{\bf Y})}~,
\end{equation}
where, crucially, ${\bf X^0}+i{\bf Y}$ and ${\bf X^0}-i{\bf Y}$ have regular OPEs.
Thus the example is essentially two copies 
of the profile (\ref{susyexp}).\footnote{The factorization could be imperfect in the
superstring case, due to the fermionic nature of the boundary fermions \cite{senspac}.
This may spoil the exact marginality of the profile.
However, we expect that it indeed factorizes since there is a path-ordered operator $P$ in the definition of the correators and this $P$ may recover the 
the marginality of the profile.}
{}From (\ref{susyBres})
we immediately find
\begin{equation}
{\cal B}(x^0,y) = {1\over 2}{\cal T}_p~{1\over |1+ 2\pi^2\lambda^2 e^{x^0+iy}|^2}~,
\end{equation}
while (\ref{susyaijres}) gives ${\cal A}^{ij} = \delta^{ij}{\cal B}$ for $i,j\neq y$, 
and (\ref{susyares}) combines with (\ref{susyBres}) to give
\begin{equation}
{\cal A}_{{1\over\sqrt{2}}(x^0+iy),{1\over\sqrt{2}}(x^0+iy)}
= {1\over 2}{\cal T}_p \left( {1\over 1+ 2\pi^2\lambda^2 e^{x^0+iy}} -2 \right)
{1\over 1+ 2\pi^2\lambda^2 e^{x^0-iy}}~.
\end{equation}
The expressions, along with the complex conjugate of the last equation, 
yields the energy momentum tensor
\begin{eqnarray}
T_{00}  &=& -{\cal T}_p ~{\rm Re} {1\over 1 + 2\pi^2\lambda^2 e^{x^0-iy}}
= - {\cal T}_p ~{1+2\pi^2 \lambda^2 e^{x^0}\cos y \over
1 + 4\pi^2 \lambda^2 e^{x^0}\cos y + 4\pi^4\lambda^4 e^{2x^0}}~,\\
T_{yy} &=& - T_{00}~,\\ 
T_{0y} &=& {\cal T}_p ~{\rm Im} {1\over 1 + 2\pi^2\lambda^2 e^{x^0-iy}}
=  {\cal T}_p ~{2\pi^2 \lambda^2 e^{x^0}\sin y \over
1 + 4\pi^2 \lambda^2 e^{x^0}\cos y + 4\pi^4\lambda^4 e^{2x^0}}~.
\end{eqnarray}
The energy momentum tensor again exhibits singularities. They appear at the 
critical time
$x^0_c = -\log(2\pi^2\lambda^2)$ and at the loci 
\begin{equation}
y_n = (2n+1)\pi , ~~~n \in Z~. 
\end{equation}
The energy density $\rho=-T_{00}$ behaves as
\begin{equation}
\rho \to {\cal T}_p 
\left[ {1\over 2}+  2\pi\sum_{n\in Z} {\rm sgn}(x^0_c-x^0)\delta(y-y_n)\right] ~,
\end{equation}
as the critical time is approached, essentially like the bosonic case (\ref{limen}). 
In the superstring case the interpretation is on a  firmer footing 
since the tachyon profile (\ref{tachssp})
amounts to the rolling down either of two sides of a symmetric, and regular, potential.
Given the topology of this situation it is not at all surprising that codimension one
defects result. Additionally, defects interpolating between the two sides of the potential
are known to couple to $RR$-fields such that, in the present case, consecutive branes 
have opposite signs, $D$-branes and anti-$D$-branes. They are sufficiently separated
that the low energy fluctuation spectrum contains no tachyons; so the configuration
is classically stable.  Nevertheless, the energy density of the defect is larger by a factor 
of $\sqrt{2}$ than that of a $BPS$ D-brane. As in the bosonic case we can verify this
result using energy conservation. 

\section{Symmetries and  Boundary States}
\label{boundary}
The purpose of this section is to reconsider the tachyon profiles in the previous 
sections using the methods and results from Sen's recent work on related 
profiles \cite{Sen1}.

\subsection{The Group of Time-Dependent Marginal Deformations}
A natural set of spatially homogeneous tachyon profiles in bosonic string theory 
is 
\begin{equation}
T(X)= \lambda_1 \cosh X^0+\lambda_2 \sinh X^0~,
\label{tg}
\end{equation}
for general $(\lambda_1,\lambda_2)$. These tachyon profiles are invariant under 
$X^0 \rightarrow X^0 +c, 
(\lambda_1 \pm\lambda_2) \rightarrow (\lambda_1 \pm \lambda_2) e^{\mp c}$;
so time translations act on the parameters $(\lambda_1,\lambda_2)$ as the 
group $SO(1,1)$. This means group invariants $\lambda_1^2-\lambda_2^2$ and 
${\rm sign}(\lambda_1 \pm \lambda_2)$ classify the possible perturbations, in the
sense that any two 
tachyon profiles of the form (\ref{tg}) with identical values of these invariants are 
physically equivalent.

As a representative of the {\it elliptic} equivalence class $\lambda_1^2-\lambda_2^2 >0$, 
we can take $T(X)=\lambda_1 \cosh X^0$. Since 
$T(X^0=0)=\lambda_1$, $\partial_0 T(X^0=0)= 0$ this corresponds to there being
a time, chosen without loss of generality as $X^0=0$, where the tachyon is displaced 
from the top of potential, but its velocity vanishes. Physically, we can thus think of the
elliptic equivalence class as having {\it negative energy}. The tachyon field starts
at the bottom of the potential, reaches a maximum at an intermediate time taken as $X^0=0$, 
and it returns to its starting point at large times.  The sign of $\lambda_1$ 
determines  which side of the potential the entire trajectory takes place, with 
positive lambda corresponding to the stable side for bosonic strings. 

As representative of the {\it hyperbolic} equivalence class $\lambda_1^2-\lambda_2^2 <0$, 
we can take $T(X)=\lambda_2 \sinh X^0$, also considered in \cite{Sen1}. For this 
profile $T(X^0=0)=0$, $\partial_0 T(X^0=0)= \lambda_2$; so this corresponds to 
there being a time where the tachyon is on top of the potential, with a non-vanishing
velocity.  Physically we can think of the hyperbolic trajectories as having {\it positive
energy}, with the tachyon field starting at the bottom on one side, reaching
the maximum the top of the potential at the time chosen as $X^0=0$, and then
rolling to the bottom of the potential on the other side. The sign of $\lambda_2$
determines which side of the potential the motion starts from.

The main focus in this paper is the {\it parabolic} equivalence class
$\lambda_1^2=\lambda_2^2$.  Taking $\lambda_1= \lambda_2(=\lambda)$ in
(\ref{tg}) gives
\begin{equation}
T(X)= \lambda e^{X^0}~.
\label{tg3}
\label{tachprof}
\end{equation}
A good physical characterization of the parabolic case is 
{\it vanishing energy}; the tachyon starts at the top of the potential, reaching 
the bottom of the potential at late times. Having no energy in the
initial state, except for the tension of the unstable brane itself, this profile 
realizes the intuition of a spontaneously decaying brane. 
Time translations can be absorbed in the magnitude of the parameter 
$\lambda$ which is thus inconsequential. Taking $\lambda_1= -\lambda_2(=\lambda)$ 
would be the time-reversed trajectory, and the sign of $\lambda$ corresponds
to the two sides of the potential.
The parabolic tachyon profile is called a half S-brane in \cite{Strominger:2002pc}, 
with the hyperbolic one being an S-brane.

The parabolic profiles can be obtained as limiting cases of the elliptic ones.
Indeed, starting from $T(X)= \lambda_1 \cosh (X^0)$ and taking the
limit  $\lambda_1 \rightarrow 0,~c \rightarrow \infty $ with fixed 
$\lambda = \frac{1}{2} \lambda_1 e^{c}$, we recover (\ref{tachprof}).
The limit corresponds to tuning the displacement of the tachyon at $X^0=0$ to 
zero, while moving the time at which the maximum is reached from 
from $X^0=0$ to the infinite past. 

Since the parabolic profiles can be represented as limits of the elliptic ones,
the corresponding energy-momentum tensors can be determined from those
computed by Sen \cite{Sen1}. For  (\ref{tg}), with $\lambda_2=0$, Sen found
the stress tensor 
$T_{00}={{\cal T}_p \over 2} (1+ \cos 2 \pi \lambda_1)$
and $T_{ij}=-\delta_{ij} T_{00} f(x^0)$,
where 
\begin{equation}
f(x^0)=\frac{1}{1+ \sin (\lambda_1 \pi) e^{x^0}} 
+\frac{1}{1+ \sin (\lambda_1 \pi) e^{-x^0}} -1~. 
\end{equation}
Generalizing this result to arbitrary elliptic $(\lambda_1,\lambda_2)$,
using the symmetry under time translation, we find
\begin{eqnarray}
T_{00}&=&{{\cal T}_p \over 2} \left[1+ \cos  (2 \pi \sqrt{\lambda_1^2-\lambda_2^2}) \right]~,\nonumber \\
 T_{ij}&=&-\delta_{ij} T_{00} f(x^0)~,
\end{eqnarray}
where 
\begin{equation}
f(x^0)=\frac{1}{1+ 
(\lambda_1+\lambda_2) 
\frac{\sin (\pi \sqrt{\lambda_1^2-\lambda_2^2})}
{\sqrt{\lambda_1^2-\lambda_2^2}  }  e^{x^0}} 
+\frac{1}{1+ 
(\lambda_1-\lambda_2) 
\frac{\sin (\pi \sqrt{\lambda_1^2-\lambda_2^2})}
{\sqrt{\lambda_1^2-\lambda_2^2}  }  e^{-x^0}} -1.
\label{emt1}
\end{equation}
Taking the limit $\lambda_1\to \lambda_2^+ (\equiv\lambda)$ to the parabolic case  
this gives $T_{00}= {\cal T}_p$ and $T_{ij}=-\delta_{ij}{\cal T}_pf(x^0)$ with
\begin{equation}
f(x^0)={1\over 1+2 \pi \lambda e^{x^0}}~, 
\label{tg2}
\end{equation}
in agreement with our explicit computations. The corresponding limit for
the superstring similarly lead to the results found in the previous section.

\subsection{Spatial Variation}
We can also consider the profile with spatial variation
\begin{equation}
T(X)=2 \lambda e^{\frac{X^0}{\sqrt{2}} }
\cos (\frac{Y}{\sqrt{2}})~,
\label{bp1}
\end{equation}
as a limit of the profile 
\begin{equation}
T(X)= \lambda_1 \cosh {\frac{X^0}{\sqrt{2}} }
\cos (\frac{Y}{\sqrt{2}})~,
\label{bp1sen}
\end{equation}
considered previously by Sen. The procedure is to replace
$X^0 \rightarrow X^0+c$ and then taking the limit
$\lambda_1 \rightarrow 0, \,\, 
c \rightarrow \infty $ with fixed
$\lambda \equiv \frac{1}{4} \lambda_1 e^{c/\sqrt{2}}$. 
The
profile (\ref{bp1sen}) was found to yield the energy momentum tensor 
\begin{eqnarray}
T_{00} &=& {\cal T}_p ~ {\rm Re} \tilde{f}~, \nonumber \\
T_{0y} &=& {\cal T}_p ~ {\rm Im} \tilde{f}~, \nonumber \\
T_{yy} &=& -{\cal T}_p ~ {\rm Re} \tilde{f}=-T_{00}~, \nonumber \\
T_{ij} &=& -{\cal T}_p ~ \delta_{ij} | \tilde{f} |^2~,
\label{tsptwo}
\end{eqnarray}
where 
\begin{equation} 
\tilde{f}=\frac{1}{1+ \sin (\tilde{\lambda} \pi/2)
\, e^{(x^0+i y)/\sqrt{2}}}
+\frac{1}{1+ \sin (\tilde{\lambda} \pi/2)
\, e^{-(x^0+i y)/\sqrt{2}}}-1~,
\end{equation}
so we can simply take the limit and find
\begin{equation}
\tilde{f}=\frac{1}{1+2\pi \lambda \, e^{(x^0+i y)/\sqrt{2}}}~.
\end{equation}
Then (\ref{tsptwo}) agrees with the results of our explicit computations (\ref{t00s}-\ref{tyys}).




\subsection{The Full Boundary States}
Since the parabolic case can be obtained as a suitable limit of other 
cases we can also use previous works to obtain the full boundary state.

Let us first review the strategy following \cite{Sen1}. 
Starting with the general profile (\ref{tg}), performing the Wick rotation $ X^0 = i X$
and redefining $\lambda_2 = - i \lambda'_2$, we find the tachyon profile 
$T(X)=\lambda_1 \cos X+\lambda'_2 \sin X$. This action contains only modes with
integer momentum modes; so we can consider the theory compactified on a self-dual 
radius $R=1$,  instead of the uncompactified theory. At this radius there is an $SU(2)$ 
current algebra with zero-modes
\begin{equation}
J^{\pm}=\oint \frac{dz}{2 \pi i} e^{\pm 2 i X_R (z)}~, \;\;
J^3=\oint \frac{dz}{2 \pi i} i \partial X_R (z)~.
\end{equation}
The tachyon profile $T(X)$ is precisely a linear combination of these  generators
and we see that the $(\lambda_1, \lambda'_2)$ can be represented as $SU(2)$
parameters as
\begin{equation}
{\cal R} = \exp \left( i \pi 
\left( \begin{array}{cc} 0 & \lambda_1+i \lambda'_2 \\ 
\lambda_1-i \lambda'_2 & 0 \end{array} \right)
\right)= 
\left( \begin{array}{cc} a & b \\ -b^* & a^* \end{array} \right),
\end{equation}
where 
\begin{equation}
a= \cos  \left(\pi \sqrt{\lambda_1^2+{\lambda'_2}^2}~\right)~, ~~~
b= i ~{\lambda_1+ i \lambda'_2\over \sqrt{\lambda_1^2+{\lambda'_2}^2}}
~\sin \left(\pi \sqrt{\lambda_1^2+{\lambda'_2}^2}~\right)
~.
\end{equation}
The boundary state for the unperturbed D-brane, with $X^0$ kept explicit,
can be written as
\cite{Callan et al}
\begin{equation}
|B \rangle_{x^0}^{\rm Neumann} =  \sum_{j=0, \frac{1}{2}, \cdots}
\sum_{m=j}^j | j,m,m \rangle \rangle ~,
\end{equation}
where $| j,m,m \rangle \rangle$ is the 
Virasoro-Ishibashi state \cite{Ishibashi} for the discrete Virasoro primary 
$|j,m,m \rangle \equiv |j,m \rangle \overline{|j,m \rangle}$. It is 
simply a sum of all Virasoro descendants of the primary $|j,m,m \rangle$.

Since the tachyon profile $T(X)$ is an element of the $SU(2)$ algebra, and 
$|j,m \rangle$ transforms in the $(j,m)$ representation of $SU(2)$ algebra, 
the non-trivial part of the boundary state becomes simply
\cite{Callan et al, PoTh, ReSc} 
\begin{equation}
|B \rangle_{x^0} =  \sum_{j=0, \frac{1}{2}, \cdots}
\sum_{m=j}^j D_{m,-m}^j ( {\cal R}) | j,m,m \rangle \rangle ~,
\end{equation}
where $D_{m,-m}^j({\cal R})$ is the spin $j$ representation matrix of 
the rotation ${\cal R}$ in $J_z$ eigenbasis (see \cite{ReSc} for an explicit form) .

To find the boundary state for the time-dependent tachyon profile (\ref{tg}), 
we then apply the appropriate inverse Wick rotation noting that, after the inverse 
Wick rotation, $b^*$ is not the complex conjugate of $b$. In the parabolic
limit $\lambda_1\to\lambda^+_2\equiv\lambda$ the "rotation" matrix becomes
\begin{equation}
{\cal R} = 
\left( \begin{array}{cc} 1 & 0 \\ 2\pi i \lambda &  1 \end{array} \right)~.
\label{parR}
\end{equation}

These considerations indeed give the correct stress tensor, for general 
$(\lambda_1,\lambda_2)$. Writing the boundary state as 
\begin{equation}
|B \rangle_{x^0} =  f(x^0) |0 \rangle +
g_{\mu \nu}(x^0) \alpha^\mu_{-1} \overline{\alpha^\nu_{-1}} |0 \rangle
+\cdots~,
\end{equation}
the stress tensor is given by 
$T_{\mu \nu}(x^0)=f (x^0)\eta_{\mu \nu} + g _{\mu \nu}(x^0)$,
as in section 2. The boundary state
$|B \rangle_{x^0} \sim \sum_{j=0, \frac{1}{2}, \cdots}
( D_{j,-j}^j | j,j,j \rangle \rangle 
+ D_{-j,j}^j | j,-j,-j \rangle \rangle)$, with
$| j,\pm j,\pm j \rangle \rangle=(i)^{2j} e^{\pm 2 i j X(0)} 
|0 \rangle+\cdots$ and $D_{j,-j}^j= (-b^*)^{2j}$,$D_{-j,j}^j=b^{2j}$
indeed lead to the $f(x^0)$ given in  (\ref{emt1}). The
$g _{\mu \nu}(x^0)$ is similarly reproduced correctly. From the boundary state point
of view the simplification offered by the parabolic case is that the
representation matrices take the simple form
\begin{equation}
D^j_{-m,m}= {(j+m)! \over (j-m)! (2 m)!  }  (2 \pi i \lambda)^{2m} \delta_{m \geq 0}~,
\end{equation}
rather than the complex formula given in \cite{ReSc}.

\section*{Acknowledgements}
We thank V. Balasubramanian, M. Einhorn, Y. He, M. Huang, T. Levi, and B. Zwiebach for 
discussions. F.L. is supported in part by DOE grant and 
A.N. is supported by DOE grant DOE-FG02-95ER40893.
A. N. and S. T. thank the Michigan Center for theoretical physics for hospitality during 
portions of this work.


\end{document}